\documentclass[conference]{IEEEtran}
\IEEEoverridecommandlockouts
\usepackage{cite}
\usepackage{url}
\usepackage[font=normalsize, labelfont=bf, skip=6pt]{caption}
\usepackage{amsmath,amssymb,amsfonts}
\usepackage[ruled]{algorithm2e}
\usepackage{graphicx}
\usepackage{textcomp}
\usepackage{placeins}
\usepackage{xcolor}
\usepackage{xcolor}
\usepackage[strings]{underscore}
\usepackage{chngcntr}
\counterwithout{table}{section}
\def\BibTeX{{\rm B\kern-.05em{\sc i\kern-.025em b}\kern-.08em
    T\kern-.1667em\lower.7ex\hbox{E}\kern-.125emX}}

\IEEEoverridecommandlockouts

\usepackage{cite}
\usepackage{amsmath,amssymb,amsfonts}
\usepackage{algorithmic}
\usepackage{graphicx}
\usepackage{textcomp}
\usepackage{xcolor}

\begin{document}

\title{eHashPipe: Lightweight Top-K and Per-PID Resource Monitoring with eBPF}
\author{
    Yuanjun Dai\textsuperscript{1,*}, Qingzhe Guo\textsuperscript{1}, Xiangren Wang\textsuperscript{2} \\
    \textsuperscript{1}Case Western Reserve University, Cleveland, OH, USA \\
    \textsuperscript{2}University of Florida, Gainesville, FL, USA \\
    \textsuperscript{*}\textit{Corresponding author: yxd429@case.edu}
}

\vspace{-12em}
\maketitle
\begin{abstract}
System-level resource monitoring with both precision and efficiency is a continuous challenge. We introduce \textbf{eHashPipe}, a lightweight, real-time resource observability system utilizing eBPF and the HashPipe sketching algorithm. eHashPipe supports two tracking modes: \emph{Top-$k$} monitoring to identify the most resource-demanding processes and specific PID tracking to detail the behavior of selected processes. We implement two in-kernel eBPF pipelines for on-CPU time and memory usage. Unlike traditional userspace polling tools, eHashPipe operates in the kernel to reduce latency and context-switch overhead while keeping the runtime footprint small. During our experiments, eHashPipe attains \textbf{100\%} \emph{Top-$k$} precision for CPU and memory at $k \in \{1, 5, 10\}$, \textbf{95.0\%/90.0\%} at $k = 20$, and \textbf{93.3\%/83.3\%} at $k = 30$ compared to the ground truth. It exposes short-lived bursts with \textbf{about $14\times$} finer temporal resolution than \texttt{top} while imposing \textbf{very low overhead}. These results show that eHashPipe delivers accurate, responsive insight with minimal impact, making it well suited for latency-sensitive cloud and edge environments.

\end{abstract}

\vspace{-0.5em}
\section{Introduction}
Modern computing systems depend on accurate and timely monitoring of system resources to ensure performance, detect anomalies, and manage workloads~\cite{rezvani2024characterizing, abel2024profiling}. Among critical resources, CPU and memory usage remain the most fundamental indicators of application health and system pressure. As systems become increasingly complex and heterogeneous, lightweight and fine-grained observability is essential not only for system administrators but also for automated runtime frameworks such as schedulers, power managers, and orchestrators~\cite{yokelson2024enabling,7920855}.

Traditional resource monitoring tools such as \texttt{top}, \texttt{pidstat}, or \texttt{vmstat} rely on periodic polling of the \texttt{/proc} interface\cite{gregg2014systems}. Although easy to deploy, these tools provide coarse-grained snapshots, especially in dynamic multiprocess workloads~\cite{vaidyanathan2006exploiting}. More advanced profiling frameworks such as \texttt{perf} or \texttt{ftrace} offer detailed tracing capabilities but incur significant runtime overhead and require expert-level tuning and interpretation~\cite{gebai2018survey}.

Recent developments in eBPF (extended Berkeley Packet Filter) have enabled a new class of low-overhead, programmable observability tools within the Linux kernel\cite{miano2023fast}. These tools support safe and dynamic instrumentation of system events, enabling insight into syscall behavior, context switches, and I/O activities~\cite{sharma2024ebpf}. However, most eBPF-based solutions focus on tracing event-level or single-resource metrics, lacking mechanisms for real-time prioritization or summarization across processes~\cite{rezvani2024characterizing}.

To address these limitations, we propose \textbf{eHashPipe}, an in-kernel resource monitoring framework that supports both \textit{Top-$k$}process summarization and targeted process tracking for CPU and memory usage. eHashPipe is based on HashPipe~\cite{sivaraman2017heavy}, a sketch-based algorithm originally developed to identify \textit{Top-$k$} flows in programmable network data planes (e.g. P4 switches and ASICs). We extend this idea from the network to the system level, designing two independent sketch-based pipelines to track CPU and memory consumption. eHashPipe maintains a dynamic summary of the most resource-intensive processes while enabling the on-demand tracking of user-specified processes via PID filtering.

Implemented entirely in the kernel using eBPF, eHashPipe eliminates the need for user-space polling and achieves a submillisecond reaction time. By placing probes at system calls and scheduling points, eHashPipe delivers high granularity and low latency observability with minimal performance overhead. Our evaluation demonstrates that eHashPipe can accurately identify the \textit{Top-$k$} CPU and memory consumers in real time while preserving modularity and extensibility for future integration of I/O or network telemetry within the same sketch-based pipeline.

\section{System Design}
Modern operating systems rely on system calls to perform essential functions such as memory allocation, I/O operations, and process scheduling~\cite{tanenbaum2015mos}. Each system call triggers a sequence of kernel-level transitions that generate a continuous stream of discrete kernel events. The eBPF framework enables fine-grained observability of these events by attaching dynamic probes (e.g., \texttt{kprobe}, \texttt{tracepoint}, \texttt{uretprobe}) at various transition points~\cite{gregg2020bpf, gebai2018survey}. Unlike network packets grouped by flow identifiers, system calls can be logically grouped into streams tied to process or thread IDs. This structural similarity motivates our use of the HashPipe via eBPF with low runtime overhead~\cite{sivaraman2017heavy}.

Our system consists of two integrated subsystems:
\begin{enumerate}
    \item A \textbf{HashPipe-based memory profiler} that tracks the usage of memory per process in real time, allowing identification of the top memory-consuming processes.
    \item A \textbf{HashPipe-based CPU utilization tracker} that captures on-CPU durations between context switch events to estimate CPU demand over time for each process.
\end{enumerate}

These components share a unified identification scheme via thread-level context (\texttt{pid}/\texttt{tgid}) and utilize lightweight BPF maps for high-performance persistence and coordination of in-kernel data.

\subsection{Memory Monitoring and Top-K Tracking via HashPipe}
Tracking memory usage in modern systems is inherently challenging due to the diversity and complexity of memory types and allocation mechanisms. Memory activity spans multiple layers, including user-level allocations, kernel-level memory management, and hardware-level intermediaries such as CPU caches and NUMA-aware memory pools. These operations are triggered through a wide range of APIs and are often influenced by the behavior of user-space libraries, runtime systems, or even compiler optimizations.

To provide a comprehensive and efficient view of memory behavior, we leverage eBPF to attach probes to critical memory-related kernel functions across both the user and the kernel space. By intercepting these events at the system or function level, we build a fine-grained memory tracking pipeline. This pipeline integrates with a lightweight HashPipe to identify \textit{Top-$k$} memory-consuming processes without requiring full state retention or complex sorting.

\subsubsection{Instrumentation and Metadata Caching}We monitor memory behavior by attaching eBPF probes to 12 key allocation functions. At function entry, we record the thread ID and allocated size in a temporary map. Subsequently, at exit, we capture the returned pointer address and store a mapping between pointer address and size for future accurate deallocation tracking. The purpose of using thread IDs instead of process IDs is to ensure precise accounting in multithreaded workloads. These events feed into a HashPipe-based pipeline, where each stage maintains fixed-size slots. Each slot holds its PID and estimated virtual memory usage, which collectively enables a fast, approximate ranking of memory-intensive processes. The following section introduces the internal logic and data update operations performed at each stage of the eHashPipe pipeline.

\begin{figure}[htbp]
    \centering
    \includegraphics[width=0.49\textwidth]{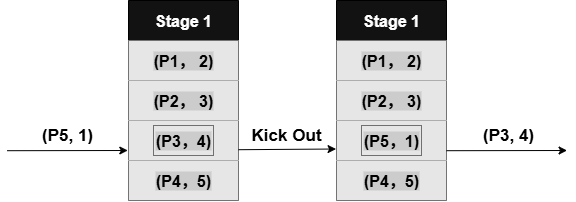}
    \caption{Always Kick Out}
    \label{fig:stage1}
\end{figure}

\begin{figure}[htbp]
    \centering
    \includegraphics[width=0.49\textwidth]{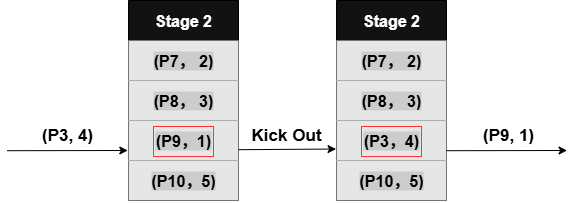}
    \caption{Comparison Kick Out}
    \label{fig:kick_out}
\end{figure}

\subsubsection{Pipeline Logic}
\begin{itemize}
    \item \textbf{Direct PID Monitoring (Lossless Tracking):} Before a memory event enters the pipeline, we first check whether its \texttt{pid} matches any user-specified target PIDs. For these selected processes, we maintain a separate BPF map that records their memory usage with full accuracy and without evictions. This lossless tracking mechanism ensures that critical or user-specified processes can be monitored precisely, regardless of their rank or competition in the main HashPipe pipeline.
    \item \textbf{Stage 0 (forced eviction):} For every memory event, a hash is computed to locate the appropriate slot in the current stage. If the \texttt{pid} matches an existing entry, its counters are updated directly. If no match is found, the algorithm checks whether the slot is empty; if so, the new entry is inserted. Otherwise, the existing entry is evicted, replaced by the new one, and forwarded to the next stage. This “always-kick” policy prevents the pipeline from stalling and ensures a continuous flow of candidate entries through the chain of stages. As we can see from Figure~\ref{fig:stage1}, under the "Always Kick Out" rule in HashPipe, the new flow (P5, 1) replaces the existing flow (P3, 4) even though it has a smaller count, ensuring that every new flow enters the pipeline.

    \item \textbf{Stage $i > 0$ (conditional replacement):} Each subsequent stage receives an entry evicted from the previous stage and processes it according to a fixed replacement strategy. If the slot is empty in the current stage, the new entry is inserted directly. If the slot is not empty but a matching \texttt{pid} is found, the corresponding counters are updated. If no matching \texttt{pid} is found and the slot is not empty, a collision occurs. The entry with the smallest \texttt{vm} is evicted, replaced by the new entry, and the evicted one is forwarded to the next stage. This cascading mechanism continues down the pipeline, enabling efficient \textit{Top-$k$} tracking while maintaining constant-time updates per stage. As we can see in Figure~\ref{fig:kick_out}, the evicted flow (P3, 4) from Stage 1 enters Stage 2 and is compared to the entry with the smallest count (P9, 1). Since P3's count 4 is larger, it replaces P9, demonstrating that kick-out decisions in later stages are based on value comparison.
\end{itemize}

\subsubsection{Memory Deallocation Process}

To track memory deallocations, the system intercepts operations such as \texttt{free}, \texttt{munmap}, and kernel-level memory deallocation. While a memory deallocation event occurs, it utilizes the pointer address to retrieve the corresponding size from the \texttt{ptr2size} table. The memory size will be sent to the eHashPipe as a negative value. The system then attempts to decrement the associated \texttt{vm} counter if a matching \texttt{pid} is found. Finally, the entry will be deleted from the \texttt{ptr2size} table to avoid stale entries.

The overall memory tracking logic, including both priority flows and approximate \textit{Top-$k$} detection, is summarized in Algorithm~\ref{alg:hashpipe_memory}.

\begin{algorithm}[!htbp]
\small
\caption{Memory Monitoring with HashPipe and Priority Flow (Kernel-Side)}
\label{alg:hashpipe_memory}
\KwIn{Memory event $e = (\texttt{pid}, \texttt{tid}, \texttt{ptr}, \texttt{size}, \texttt{type})$}
\KwOut{Updated memory entries}

// Step 1: Handle user-specific PID with exact tracking\;
\If{$e.\texttt{pid} \in \texttt{SpecifiedPIDs}$}{
    \If{$e.\texttt{type} == \texttt{allocation}$}{
        $\texttt{priority\_vm}[e.\texttt{pid}] += \texttt{size}$\    }
    \ElseIf{$e.\texttt{type} == \texttt{deallocation}$}{
        $\texttt{size} \leftarrow \texttt{ptr2size}[e.\texttt{ptr}]$\;
        delete $\texttt{ptr2size}[e.\texttt{ptr}]$\;
        $\texttt{priority\_vm}[e.\texttt{pid}] -= \texttt{size}$\;
    }
    \Return\;
}

\If{$e.\texttt{type} == \texttt{allocation}$}{
    $\texttt{ptr2size}[e.\texttt{ptr}] \leftarrow \texttt{size}$\;
    $entry \leftarrow (e.\texttt{pid}, \texttt{size})$\;
}
\ElseIf{$e.\texttt{type} == \texttt{deallocation}$}{
    \If{$e.\texttt{ptr} \in \texttt{ptr2size}$}{
        $size \leftarrow \texttt{ptr2size}[e.\texttt{ptr}]$\;
        delete $\texttt{ptr2size}[e.\texttt{ptr}]$\;
        $entry \leftarrow (e.\texttt{pid}, -size)$\;
    }
    \Else{
        \Return\;
        }
}
 \For{$i = 0$ \KwTo $d-1$}{
    $h_i \leftarrow (a_i \cdot e.\texttt{pid} + b_i) \bmod N$\;

    \If{slot $S_i[h_i]$ is empty}{
        $S_i[h_i] \leftarrow entry$;
        \Return\;
    }
    \ElseIf{$S_i[h_i].pid == entry.pid$}{
        $S_i[h_i].vm \leftarrow S_i[h_i].vm + entry.size$\;
        \Return\;
    }

    \If{$i == 0$}{
        \tcp{Stage 1: Always kick out}
        swap($entry$, $S_0[h_0]$)\;
    }
    \ElseIf{$entry.vm > S_i[h_i].vm$}{
        \tcp{Stage $i \geq 1$: Compare and swap}
        swap($entry$, $S_i[h_i]$)\;
    }
    \Else{
        \Return\;
    }
    \Return\;
    }
\end{algorithm}

\subsection{Fine-Grained On-CPU Time Tracking via HashPipe}

To capture real-time CPU utilization patterns per process, we implement an on-CPU time tracking system using eBPF, built on top of the \texttt{sched\_switch} tracepoint. This tracepoint is triggered by the Linux kernel scheduler every time the CPU context switches from one thread to another. Importantly, \texttt{sched\_switch} operates at the granularity of \textbf{threads} rather than processes. Each scheduling event provides access to the \texttt{ task_struct} of the thread being switched out (\texttt{prev}), while the thread being scheduled in can be identified by \texttt{bpf_get_current_pid_tgid()}. We can obtain both the thread ID (TID) and the process ID (PID) from this function.

Our implementation records the timestamp for each scheduled thread and stores it in a global \texttt{BPF_HASH} map keyed by thread ID. When the same thread is later scheduled out in a future invocation of \texttt{sched\_switch}, we retrieve its entry, compute the delta time, and attribute it to the main process. This delta reflects the precise duration the thread spent on-CPU during its most recent scheduling cycle.

The eHashPipe used for on-CPU time tracking shares the same architecture as the memory tracking pipeline. The primary difference lies in the update semantics. While memory tracking requires both addition (for allocations) and subtraction (for deallocations), on-CPU time tracking is inherently monotonic. As a result, the on-CPU pipeline only performs additive updates. This simplifies both the implementation and run-time behavior.

Similarly, we use a dedicated BPF map to monitor a user-specified process. This per-thread map provides fine-grained observability without approximation and is designed to track critical or user-defined workloads.

Together, HashPipe-based approximate aggregation and dedicated per-process tracking offer a hybrid observability mechanism: scalable approximated summarization at the system level and lossless monitoring for user-defined targets.

\subsection{Summary}
Our system uses two HashPipe-based pipelines, one for memory and one for CPU, to track \textit{Top-$k$} resource-intensive processes with low overhead. The memory pipeline monitors allocation and deallocation events, while the CPU pipeline records on-CPU time. For user-defined processes, we provide dedicated maps for precise full-resolution tracking along the approximate \textit{Top-$k$} pipeline.

\section{Implementation}

Our eBPF-based resource tracking system incorporates a HashPipe-inspired structure to efficiently monitor per-process CPU and memory usage. The design emphasizes configurability, loop-based pipelining, and robustness under high-frequency system events. We highlight four core considerations in our implementation.

\paragraph{Configurability}
To support diverse workloads and system scales, our implementation allows users to configure both the number of pipeline stages and the number of slots per stage. However, due to strict verifier constraints in eBPF, such as limited stack size, instruction count limits, loop unrolling restrictions, and overall code size caps, the configuration space must be carefully tuned. An excess stage count or map size may lead to verifier rejection or program loading failure. In practice, we empirically determine the safe ranges for our setup.

\paragraph{Physically Unrolled Loops}

Our HashPipe implementation models each event as passing through a fixed number of pipeline stages. Since eBPF requires all control flow to be statically analyzable, we cannot use traditional run-time loops. To address this, we use the \texttt{\#pragma clang loop unroll(full)} directive to fully unroll the pipeline loop at the compile time\cite{clangpragma}. This transforms the loop into a linear sequence of instructions, eliminating data-dependent branches and ensuring compliance with the eBPF verifier.

\paragraph{Data Consistency}

Although our system tracks memory events at the thread level to reduce key collisions, we still observe race conditions under high-intensity workloads. This issue is especially common in C++-based machine learning frameworks, which frequently allocate and deallocate complex data structures, resulting in a high volume of memory events. To ensure consistency, we use atomic operations \texttt{__sync\_fetch\_and\_add()} and \texttt{__sync\_fetch\_and\_sub()} to update the table entry safely.

\section{Evaluation}

We design an evaluation framework to assess the effectiveness, precision, and efficiency of our eBPF-based monitoring system. Our evaluation consists of two key dimensions: (1) tracking accuracy for top-$k$ process identification, and (2) system overhead including memory footprint and CPU impact.

\subsection{Metrics}
\begin{itemize}
\item \textbf{Top-k Fidelity:} Accuracy of \textit{Top-$k$} detection.
\item \textbf{Memory Overhead:} Memory used to track event flows.
\item \textbf{CPU Overhead:} Runtime overhead.
\item \textbf{Responsiveness:} Frequency to update \textit{Top-$k$} list.

\end{itemize}
\subsection{Experimental Setup}
All experiments were conducted on a commodity laptop running Ubuntu 22.04. The system is equipped with a 6-core, 12-thread x86\_64 CPU with a maximum clock speed of 4.3\,GHz, 16\,GB of RAM, and Linux kernel version 5.15.167.
\subsection{Workloads and Ground Truth Construction}
Our evaluation includes both synthetic and real-world workloads. For synthetic testing, we simulate high-frequency memory allocations and context switches using fork bombs. These workloads are designed to test the robustness and accuracy of our system under controlled stress conditions.

For real-world testing, we monitor commonly used service daemons such as nginx, as well as CPU-intensive training tasks based on PyTorch. These workloads introduce realistic, dynamic resource usage patterns, enabling us to assess the responsiveness of the system in practical deployment scenarios.

To construct a ground truth, we capture real-time snapshots of the top consumers of the system using the top command, which aggregates values directly from the \texttt{ /procs} file system. This approach ensures a reliable baseline for comparing the output of eHashPipe.
\subsection{Top-k Accuracy Evaluation}

To evaluate the fidelity of our sketch-based \textit{Top-$k$} summarization, we assess the precision of eHashPipe in identifying the main processes that consume resources with varying values of $k$. We consider both CPU and memory usage and compare the output of eHashPipe against ground truth. The detection frequency of eHashPipe is configured once every 2 seconds, aligned with the sampling interval of \texttt{top} to ensure a fair comparison.
\begin{table}[ht]
\centering
\begin{tabular}{|c|c|c|}
\hline
\textbf{Top-$k$} & \textbf{CPU Accuracy (\%)} & \textbf{Memory Accuracy (\%)} \\
\hline
1   & 100.0 & 100.0 \\
5   & 100.0  & 100.0  \\
10  & 100.0  & 90.0  \\
20  & 95.0 & 90.0  \\
30  & 93.3  & 83.3  \\
\hline
\end{tabular}
\vspace{0.5em}
\caption{Top-$k$ Accuracy of eHashPipe vs. Ground Truth}\label{tab:topk_accuracy}
\end{table}
We define \textit{Top-$k$} accuracy as the percentage of overlap between the set of \textit{Top-$k$} PIDs reported by eHashPipe and the corresponding \textit{Top-$k$} list extracted from \texttt{top} at the same time stamp. Table~\ref{tab:topk_accuracy} summarizes the accuracy values for $k \in \{1, 5, 10, 20, 30\}$ in both CPU and memory monitoring. Although both methods achieve perfect accuracy at $k=1$, eHashPipe maintains over 85\% accuracy at $k=30$, validating its effectiveness even under relaxed detection thresholds.

\subsection{Responsiveness}
To evaluate the temporal sensitivity of system monitoring tools, we conducted an experiment comparing our proposed eHashpipe with the traditional user-space \texttt{top} command. The target workload involves training a ResNet-34 model on the CIFAR-10 dataset using PyTorch on a CPU-only system. During the training process, we monitored both CPU and memory usage using both tools.

Although the \texttt{top} command allows users to specify fine-grained sampling intervals (e.g., 0.001s or 0.01s), our measurements indicate that its effective sampling resolution saturates around 0.15 seconds. Consequently, even under its most aggressive configuration, \texttt{top} fails to capture short-lived fluctuations and transient bursts in system behavior.

\begin{figure}[h]
  \centering
  \includegraphics[width=0.75\linewidth]{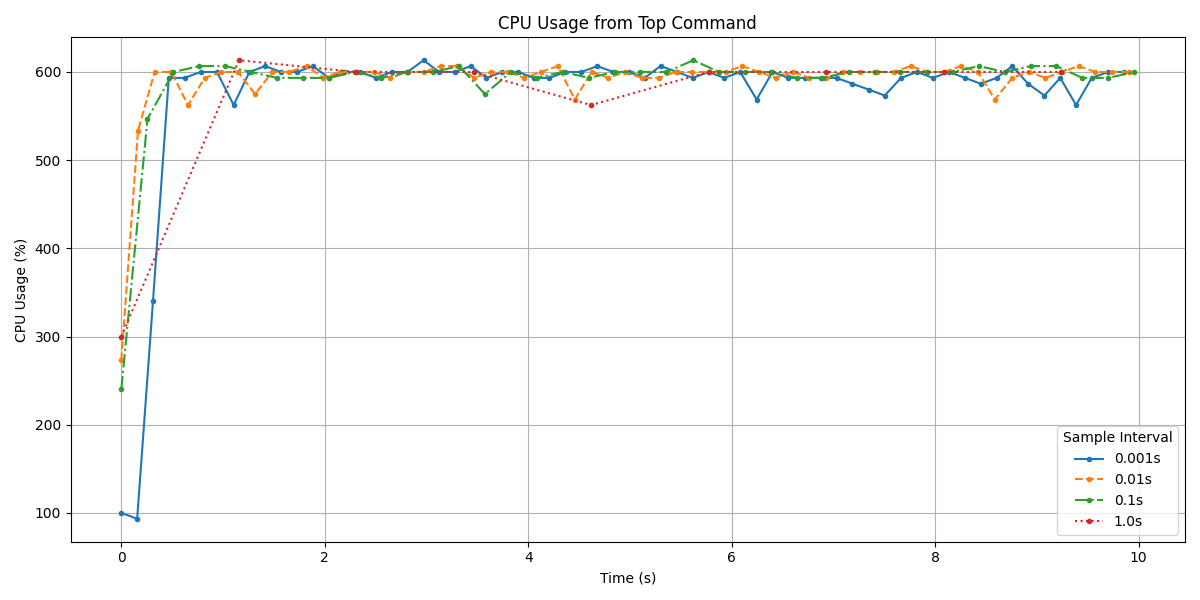}
  \caption{CPU usage recorded by top under various intervals.}
  \label{fig:cpu-top}
\end{figure}

\begin{figure}[h]
  \centering
  \includegraphics[width=0.75\linewidth]{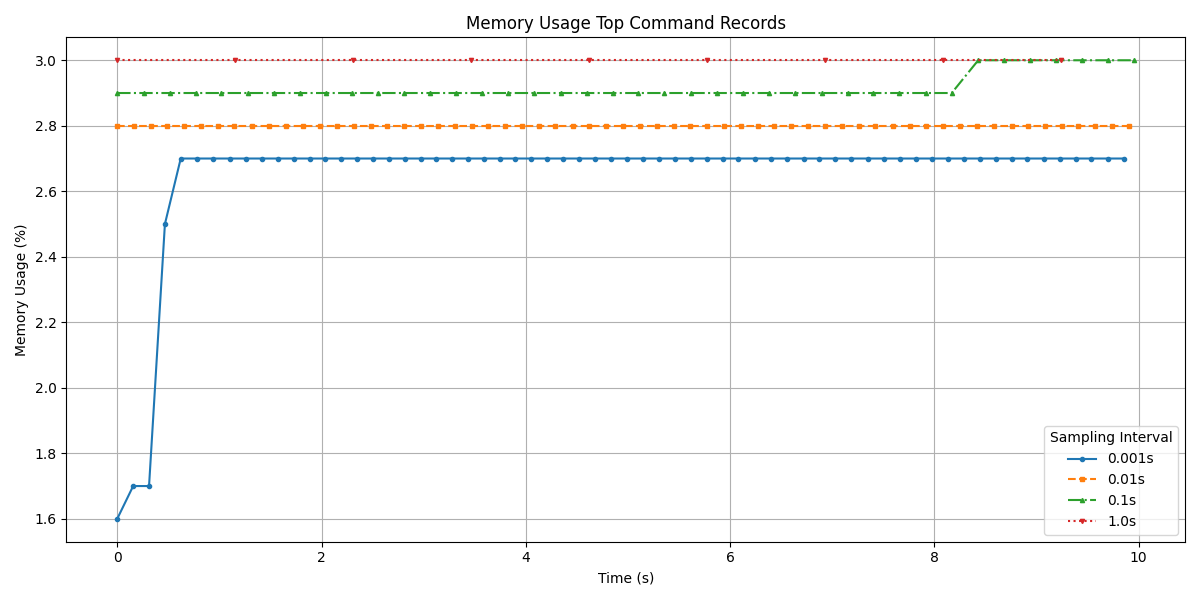}
  \caption{Memory usage recorded by top under various intervals.}
  \label{fig:mem-top}
\end{figure}

\begin{figure}[h]
  \centering
  \includegraphics[width=0.75\linewidth]{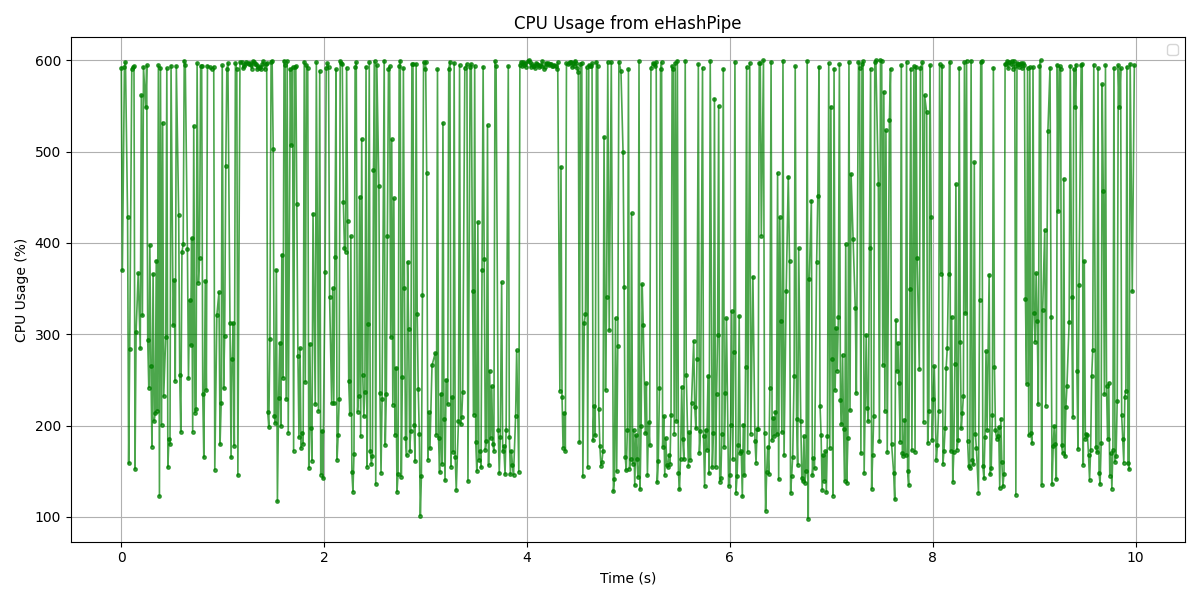}
  \caption{CPU usage recorded by eHashpipe.}
  \label{fig:cpu-ehashpipe}
\end{figure}

\begin{figure}[h]
  \centering
  \includegraphics[width=0.75\linewidth]{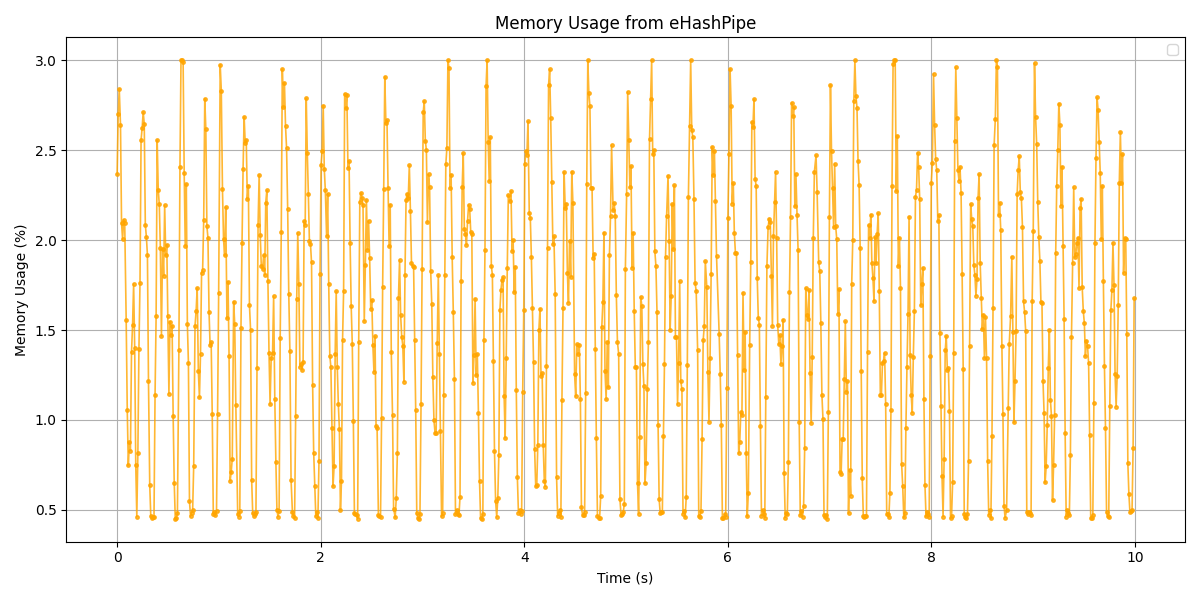}
  \caption{Memory usage recorded by eHashpipe.}
  \label{fig:mem-ehashpipe}
\end{figure}

We visualize these differences in Figures~\ref{fig:cpu-top}–\ref{fig:mem-ehashpipe}. Figures~\ref{fig:cpu-top} and~\ref{fig:mem-top} show CPU and memory usage recorded by \texttt{top} under various sampling intervals. Regardless of the configuration, the results appear coarse and flattened, missing critical temporal variations. In contrast, eHashpipe operates with a target sampling interval of 0.01 seconds and achieves an effective temporal resolution of approximately 0.011 seconds—an order of magnitude finer than \texttt{top}.

This discrepancy becomes particularly significant during deep learning training workloads, which are composed of repeating iterations involving multiple phases: forward pass, backward pass, data loading and release, and intermediate tensor allocation and deallocation. These phases alternate between CPU-intensive and memory-intensive operations and generate substantial memory churn. Due to the repetitive nature of training iterations, both CPU and memory usage exhibit cyclic patterns\cite{dai2023dnn}. However, \texttt{top}'s limited temporal resolution and smoothing effects obscure these periodic behaviors, failing to reflect the underlying dynamics of the target application.

Figures~\ref{fig:cpu-ehashpipe} and~\ref{fig:mem-ehashpipe} illustrate the fine-grained traces captured by eHashpipe. The results show clear high-frequency oscillations in both CPU and memory usage. Notably, memory usage fluctuates periodically between 0.5\% and 3.0\%, highlighting micro-level transitions that are invisible to \texttt{top}. Similarly, CPU usage traces reveal bursty and phase-specific patterns aligned with training logic.

In summary, eHashpipe significantly enhances the temporal granularity and responsiveness of system monitoring. It enables accurate profiling of short-lived workloads, detection of transient spikes, and precise diagnosis of system performance during complex operations like deep model training.

\subsection{Resource Overhead}

In our evaluation setup, we measured the memory footprint and CPU overhead of both HashPipe modules (memory and CPU).

\textbf{Memory usage.} The eHashPipe memory module uses approximately 1.2\,MB, including 5 inner maps (2000 entries each, 40 bytes per entry) and auxiliary maps. The CPU HashPipe module uses about 183\,KB.

\textbf{CPU Overhead.}
Measured with \texttt{perf stat}, the eBPF process incurs only about \textbf{20\%} CPU overhead under our evaluation setup.

\subsection{Summary of Findings}

Our evaluation demonstrates that eHashpipe provides a lightweight, high-fidelity, and responsive in-kernel monitoring solution that significantly outperforms traditional user-space tools such as \texttt{top} in several key aspects. First, in terms of tracking accuracy, eHashpipe maintains over 80\% top-$k$ fidelity for both CPU and memory resource identification, even as $k$ increases. Second, in terms of temporal sensitivity, it captures fine-grained system snapshots and is orders of magnitude more responsive than \texttt{top}. Third, despite operating at high frequency, the system maintains an acceptable resource overhead.

Together, these results establish eHashpipe as a flexible and accurate kernel-level telemetry mechanism for capturing short-lived, high-volume, or bursty behaviors. Its accuracy, responsiveness, and modular design make it well-suited for use in cloud observability platforms, edge performance analytics, and runtime diagnostics of latency-sensitive applications.

\section{Conclusion and Future Work}
In this paper, we introduce a lightweight and scalable system for real-time monitoring of memory and CPU usage using eBPF and a HashPipe-based  \textit{Top-$k$} algorithm. Our design offers configurable, low-overhead tracking and performs well across diverse workloads. In addition to approximate \textit{Top-$k$} detection, the system also supports accurate monitoring of user-specified processes.

Future work will focus on developing an integrated performance monitoring system that incorporates additional system parameters---particularly those relevant to distributed machine learning (DML) workloads, such as GPU utilization, memory bandwidth, and inter-node communication latency. By combining our eHashPipe framework with fine-grained GPU and I/O metrics, we aim to build a holistic tool for profiling and diagnosing performance bottlenecks across the entire DML training pipeline\cite{jia2019flexflow,pan2022efficient}.
Building upon this monitoring infrastructure, we plan to enable real-time performance tuning and resource optimization for DML tasks, leveraging dynamic insights to guide adaptive scheduling, batch size scaling, and communication optimization.

Finally, the underlying concepts behind eHashPipe, especially its sketch-based low-overhead telemetry, are equally applicable to user-level applications. One promising direction is to apply this approach to electronic health record (EHR) systems, where large-scale, high-frequency data processing demands efficient and scalable analytics\cite{volk2014development}. Our techniques can help optimize health service applications by improving observability and reducing computational overhead in data-intensive healthcare environments.
\bibliographystyle{IEEEtran}
\bibliography{ref}
\end{document}